\documentclass[12pt]{article}

\usepackage[T1]{fontenc}
\usepackage[utf8]{inputenc}
\usepackage{times}
\usepackage{geometry}
\usepackage{setspace}
\usepackage{physics}
\usepackage{graphicx}
\usepackage{amssymb}
\usepackage{amsmath}
\usepackage{dsfont}
\usepackage{bm}
\usepackage{mathrsfs}
\usepackage{bbold}
\usepackage{xcolor}
\usepackage{amsthm}
\usepackage{nameref}
\usepackage{hyperref}

\hypersetup{
  colorlinks=true,
  linkcolor=blue,
  citecolor=blue,
  urlcolor=blue
}

\newtheoremstyle{natural}%
{\topsep}{\topsep}%
{\normalfont}{}%
{\normalfont}{.}{5pt}%
{}
\theoremstyle{natural}

\begin{document}

\begin{center}
{\Large \textbf{Singularities as Solitons?}\\[-0.2em]{\small Quantum Vacuum Architecture of Black Holes}}\\[1em]

Eren Erberk Erkul$^{|1\rangle,|2\rangle}$\\
{\small
$^{|1\rangle}$Institute of Science and Technology Austria (ISTA), 3400 Klosterneuburg, Austria\\
$^{|2\rangle}$Department of Physics \& Department of Electrical and Electronics Engineering,\\
Middle East Technical University (METU), Ankara 06800, Turkey
}\\[0.5em]
\textbf{Dated:} October 9, 2025
\end{center}

\begin{center}
\textbf{Abstract}
\end{center}

\begin{center}
\begin{minipage}{0.9\textwidth}
We propose that black holes are \emph{soliton-esque} objects, where gravitational collapse is balanced by quantum
vacuum dispersion, modeled via \(R+\alpha R^{2}\) gravity. Classical singularities are replaced by oscillating,
finite-radius cores, thereby evading static no-go theorems. The event horizon is replaced by the \textit{Lamarina},
a surface of maximum redshift whose surface geometry yields Hawking-like radiation with corrections. The Raychaudhuri equations
impose a Dyson-type ceiling on the maximum radiated power \((P_{\infty} \lesssim c^{5}/G)\), while effective field theory
matching dictates a universal minimum Lamarina radius set by the dispersion scale.
\end{minipage}
\end{center}

\newpage

{\small\emph{Conventions.} Signature $(-,+,+,+)$. Geometric units $G=c=1$ are used except where $c$ is explicitly restored in flux/power formulas.}

A simple dimensional estimate first done by Freeman Dyson \cite{Dyson1963,JowseyVisser2021} suggests a fundamental upper limit on the power of gravitational radiation. Exceeding this scale would concentrate energy so intensely that horizon-like regions should form and cut off further emission. Yet classical general relativity neither enforces this bound nor predicts such a mechanism. Numerical relativity simulations \cite{Cardoso2018,Sperhake2008,Keitel2017} and LIGO results \cite{Abbott2016Properties}, however, indicate that in highly non-linear regimes, horizons do form, effectively self-limiting radiation. Classically, bounds on radiative losses can be formulated via the Bondi mass-loss rate \cite{CaoLiWu2021}, but these do not by themselves enforce a universal, source-independent ceiling.

We propose that the missing ingredient in classical general relativity to capture the idea of the power ceiling is the dispersive nature of the quantum vacuum \cite{ErkulLeonhardt2025}. Rather than a perfectly smooth manifold, spacetime behaves more like rugged terrain: near extreme curvatures, its granular structure becomes dominant. This dispersion counteracts the runaway collapse of gravity, in direct analogy to how solitons arise when dispersion balances nonlinearity. In our view, black holes do not collapse to true singularities. Instead, they settle into stable, soliton-like configurations, bounded by a membrane that replaces the classical horizon \cite{MazurMottola2004,HoldomRen2017,Holdom2022}.\footnote{Soliton structures also appear within classical general relativity itself through solution-generating techniques \cite{NeugebauerKramer1981}, but here we argue for a fundamentally quantum-vacuum origin encoded by a minimal curvature correction.} More broadly, the configuration we discuss fits within the wider landscape of horizonless compact-object models and their observational probes \cite{CardosoPani2019}.

\noindent\textit{\textbf{Quantum Vacuum Soliton.}}---
The intrinsic nature of quantum-vacuum forces is fundamentally incompatible with classical general relativity. Nevertheless, semiclassical insights have resulted in major breakthroughs such as Hawking radiation~\cite{Hawking1974}, the Gibbons--Hawking effect~\cite{GibbonsHawking1977}, and the Unruh effect~\cite{Unruh1976}. Conceptually, dispersion is expected to arise from even-powered derivative corrections~\cite{Runge2021}; in gravity, the leading such correction is quadratic in curvature~\cite{tHooftVeltman1974,Starobinsky1980}.

Guided by this, and to avoid the spin-2 ghost of generic quadratic gravity \cite{Stelle1977}, we adopt the classical, ghost-free metric $f(R)$ model
\begin{equation}
S=\frac{1}{16\pi G}\int d^4x\,\sqrt{-g}\,\big(R+\alpha R^2\big),\qquad \alpha>0.
\label{eq:action}
\end{equation}
The field equations are
\begin{equation}
(1+2\alpha R) R_{\mu\nu}-\tfrac12 (R+\alpha R^2) g_{\mu\nu}
-\big(\nabla_\mu\nabla_\nu-g_{\mu\nu}\Box\big)(1+2\alpha R)=0.
\label{eq:fR_eom}
\end{equation}
This theory is dynamically equivalent to GR plus one healthy scalar degree of freedom (the scalaron) with mass
\begin{equation}
m_\varphi^2=\frac{1}{6\alpha},
\qquad L_{\rm QV}:=m_\varphi^{-1}=\sqrt{6\alpha},
\label{eq:scalaron}
\end{equation}
which sets the microscopic dispersion length and the proper thickness of the wall. We assume the usual viability conditions $f_R=1+2\alpha R>0$ and $f_{RR}=2\alpha>0$ in the region of interest \cite{SotiriouFaraoni2010,DeFeliceTsujikawa2010}. We treat $\alpha$ as a phenomenological Wilson coefficient capturing short-distance quantum-vacuum dispersion. This is consistent with the standard view of GR as a low-energy effective field theory \cite{Donoghue1994}. Hence, our results depend only on the sign of $\alpha$ and the length $L_{\rm QV}=\sqrt{6\alpha}$.

To investigate the structure of the proposed soliton, we use a quasi-stationary (adiabatic) metric:
\begin{equation}
ds^2 = -A(t,r)\,dt^2 + B(t,r)\,dr^2 + r^2\big(d\theta^2 + \sin^2\theta\, d\phi^2\big),
\label{eq:ansatz}
\end{equation}
with $A,B>0$. The adiabatic parameter compares background drift to the local wall scale,
\begin{equation}
\varepsilon \sim
\max\left\{
\frac{\dot A}{A \,\kappa_{\rm loc}},
\frac{\dot B}{B \,\kappa_{\rm loc}},
\frac{\dot R_L}{R_L \,\kappa_{\rm loc}}
\right\} \ll 1.
\label{eq:adiabaticity_parameter}
\end{equation}
Regularity and isotropy at the center ($r=0$) on any fixed time slice require even-power expansions with a curved but regular center:
\begin{subequations}\label{eq:series_expansion}
\begin{align}
A(t,r) &= 1 - a_2(t)\, r^2 + a_4(t)\, r^4 + a_6(t)\, r^6 + \mathcal{O}(r^8), \label{eq:series_expansion_a} \\[6pt]
B^{-1}(t,r) &= 1 - b_2(t)\, r^2 + b_4(t)\, r^4 + b_6(t)\, r^6 + \mathcal{O}(r^8). \label{eq:series_expansion_b}
\end{align}
\end{subequations}
so the central curvature is
\begin{equation}
R(t,0)\equiv R_c(t)=6\,[a_2(t)+b_2(t)].
\label{eq:R0def}
\end{equation}
Because \eqref{eq:fR_eom} is fourth order, the leading center equations couple to next orders: the trace of \eqref{eq:fR_eom} yields the massive Klein-Gordon equation \cite{SotiriouFaraoni2010,DeFeliceTsujikawa2010}
\begin{equation}
\Box R - m_\varphi^2\,R=0,
\label{eq:trace-eq}
\end{equation}
which near the center fixes the next Taylor coefficient instantaneously. Approximating the d'Alembertian by $\Box R(t,0)\simeq -\partial_t^2 R_c(t)+6R_2(t)$ one obtains
\begin{equation}
6R_2(t)=\big(m_\varphi^2+\partial_t^2\big)R_c(t).
\label{eq:Rseries}
\end{equation}
For a time-harmonic core $R_c(t)=\Re\{R_{c0}e^{i\Omega t}\}$ this gives
\begin{equation}
6R_2(t)=(m_\varphi^2-\Omega^2)\,R_c(t).
\label{eq:Rseries-harmonic}
\end{equation}
Linearizing about a weak-field background and taking plane waves $e^{-i\omega t+i\mathbf{k}\cdot\mathbf{x}}$ yields the dispersion relations
\begin{equation}
\omega^2=k^2 \quad \text{(tensor)},\qquad \omega^2=k^2+m_\varphi^2 \quad \text{(scalaron)},
\label{eq:dispersion}
\end{equation}
so the two GR tensor polarizations remain luminal and non-dispersive, while the scalar degree of freedom is dispersive with group velocity $v_g=\partial\omega/\partial k=\sqrt{1-m_\varphi^2/\omega^2}$. For sub-threshold local frequencies $\omega<m_\varphi$ one has $k=i\kappa$ with $\kappa=\sqrt{m_\varphi^2-\omega^2}$.

\noindent\textit{\textbf{Static no-go in $R+\alpha R^2$ vacuum.}}---
Consider a static, spherically symmetric, horizonless, and asymptotically flat vacuum configuration in the model $f(R)=R+\alpha R^2$ with $\alpha>0$. 

The trace of the field equations is a massive Klein–Gordon equation for the Ricci scalar,
\begin{equation}\label{eq:trace-KG}
(\Box - m_\varphi^{2})\,R = 0,
\qquad
m_\varphi^{2}=\frac{1}{6\alpha}>0.
\end{equation}
In a static setting, multiplying this equation by $R$ and integrating over $r$ results in an identity,
\begin{equation}\label{eq:energy-identity}
\begin{aligned}
&\big[r^{2}\sqrt{A/B}\,R\,R'\big]_{0}^{\infty}\\
&= \int_{0}^{\infty}\!dr\,\Big(r^{2}\sqrt{A/B}\,(R')^{2}+r^{2}\sqrt{AB}\,m_\varphi^{2}R^{2}\Big)\\
&= 0.
\end{aligned}
\end{equation}
whose boundary term vanishes by regularity at the center and decays at infinity. Both sides are nonnegative and thus must vanish, forcing $R\equiv 0$.

The vacuum equations then reduce to $R_{\mu\nu}=0$, i.e., the Schwarzschild family; with “horizonless” imposed, only the $M=0$ member (Minkowski) remains. Thus any nontrivial vacuum configuration in $R+\alpha R^2$ must be \emph{time dependent} (cf.\ Jebsen--Birkhoff in $f(R)$ \cite{Faraoni2010}).

Henceforth we work with a sub-threshold, time-periodic scalaron core (an adiabatic ``oscillaton'') \cite{SeidelSuen1991} to evade the static no-go while keeping horizonlessness and asymptotic flatness. Localized self-gravitating scalar configurations have a long history \cite{Kaup1968,RuffiniBonazzola1969}; see \cite{LieblingPalenzuela2023} for a modern review:
\begin{equation}
R(t,r) = \mathcal{R}(r)\cos(\Omega t), \qquad 0 < \Omega < m_\varphi,
\label{eq:oscillaton_ansatz}
\end{equation}
Far from the wall, where $A(r)\to 1$ and $|R|\ll m_\varphi^2$, the linearized trace equation reduces to $(\nabla^2-\kappa^2)\mathcal R=0$ with
\begin{equation}
\mathcal R(r)\propto \frac{e^{-\kappa r}}{r},\qquad \kappa=\sqrt{m_\varphi^2-\Omega^2}>0,
\label{eq:exterior-tail}
\end{equation}
which ensures decay at infinity. This tail is an exterior solution only; regularity at the origin is secured by the interior region and global matching across the oscillatory shell. By contrast, in flat space the unique regular solution at the origin behaves like $\sinh(\kappa r)/r$ and grows, illustrating why curvature/redshift is essential for localization.

A necessary condition for localized time-harmonic profiles is obtained as follows. For the time-harmonic ansatz in a static SSS background, the radial equation can be written as
\begin{equation}
\begin{aligned}
\partial_r\!\big(p\,\psi'\big) &= w\,\Big(m_\varphi^2-\frac{\Omega^2}{A(r)}\Big)\psi,\quad
p:=r^2\sqrt{\frac{A}{B}}>0,\\
w &:= r^2\sqrt{AB}>0
\end{aligned}
\label{eq:radial-identity}
\end{equation}
where $\psi(r):=\mathcal R(r)$. Multiplying by $\psi$ and integrating from $0$ to $\infty$ (regular center and decaying tail annihilate the boundary term) gives
\begin{equation}
-\int_0^\infty p(\psi')^2\,dr=\int_0^\infty w\Big(m_\varphi^2-\frac{\Omega^2}{A(r)}\Big)\psi^2\,dr.
\label{eq:integral-identity}
\end{equation}
A necessary condition for a nontrivial solution is that the integrand on the right change sign somewhere, i.e.\ there exists a region where $m_\varphi^2-\Omega^2/A(r)<0$. Since $A(r)\ge A_{\min}$, this requires the frequency window
\begin{equation}
m_\varphi\sqrt{A_{\min}} < \Omega < m_\varphi.
\label{eq:freq-window}
\end{equation}
In ultra-compact geometries with $A_{\min}\ll 1$ this window is nonempty, allowing a localized profile concentrated near the wall.

\noindent\textit{\textbf{The Lamarina and Emission.}}---
Rather than forming a singularity at $A=0$ shielded by a horizon in the Schwarzschild solution, the spacetime settles into a regular configuration where the lapse $A(t,r)$ remains globally positive and develops a sharp minimum.

\begin{figure}[t!]
  \centering
  \hspace*{19mm} 
  \includegraphics[width=0.9\textwidth]{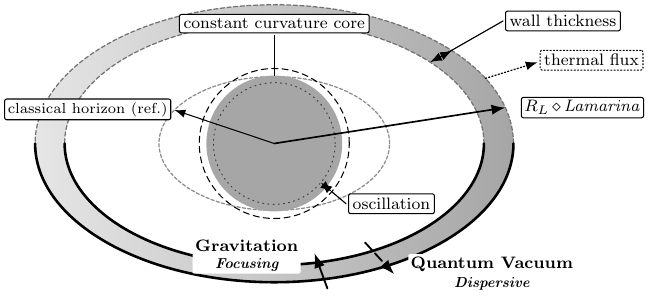}
  \caption{Schematic representation of the proposed quantum vacuum soliton structure. Illustrating the \textit{Lamarina} replacing the classical horizon and the localized curvature profile.}
  \label{fig:soliton_schema}
\end{figure}

This minimum corresponds to the radius where the gravitational pull is balanced by dispersion. We term this surface the \textbf{\textit{Lamarina}} \footnote{Turkish Cypriot term of Latin origin meaning metal covering} and denote its instantaneous radius by $R_{L}(t)$. This replaces the classical horizon and defines the surface of maximum gravitational redshift:
\begin{equation}
\partial_r A\big(t,R_{L}(t)\big)=0,\quad A\big(t,R_{L}(t)\big)=:A_{\min}(t)>0.
\label{eq:lamarina_def}
\end{equation}
Writing $y(t,r)=\ln A(t,r)$ and $dx=\sqrt{B}\,dr$ gives the universal quadratic approximation in proper distance near the minimum (instantaneously),
\begin{equation}
y(t,x) \approx y_0(t) + \frac{1}{2}\kappa_{\text{loc}}(t)^2 x^2 + \mathcal{O}(x^3),
\label{eq:quadratic_expansion}
\end{equation}
with $y_0(t)=\ln A_{\min}(t)$ and local curvature scale
\begin{equation}
\kappa_{\text{loc}}(t)^2 = \frac{d^2y}{dx^2}\bigg|_{r=R_{L}(t)}.
\label{eq:kappa_def}
\end{equation}
Casting \eqref{eq:fR_eom} in Einstein form $G^\mu{}_\nu=8\pi G\,T^{\rm eff\,\mu}{}_\nu$ with the standard $f(R)$ effective stress, one finds (details in Supplemental Material) the wall identities; in particular, instantaneously,
\begin{equation}
\kappa_{\text{loc}}^2 = 8\pi G\big[\rho_{\rm eff}+p_{r,{\rm eff}}+2p_{t,{\rm eff}}\big]\Big|_{r=R_{L}(t)}>0,
\label{eq:k_loc_identity}
\end{equation}
so the minimum is supported by the $R^2$ sector. Because ultra-compact horizonless configurations typically feature light rings and long-lived trapped modes, they can be prone to instabilities; whether the scalaron sector and Lamarina boundary conditions suppress those mechanisms is an open question \cite{CardosoUCOInstability2014,Keir2016}.

With this local form in hand, null geodesics skimming the quadratic minimum of $y=\ln A$ at $r=R_L$ exhibit an exponential peeling map (adiabatic regime), yielding a universal Planck factor with effective temperature
\begin{equation}
T_{\rm eff}=\frac{\hbar}{2\pi k_B}\,\kappa_{\rm peel},
\label{eq:T_eff}
\end{equation}
and Tolman redshift to infinity (adiabatic) \cite{TolmanEhrenfest1930},
\begin{equation}
T_\infty=\sqrt{A_{\min}}\,T_{\rm eff}.
\label{eq:T_infty}
\end{equation}
Here $\kappa_{\rm peel}$ is the peeling rate; in the strict adiabatic limit $\kappa_{\rm peel}=\kappa_{\rm loc}+\mathcal O(\varepsilon)$ (derivation in Supplement~B).

\noindent\textit{\textbf{The Dyson Ceiling.}}---
Beyond thermal emission, regularity of the core implies a fundamental constraint on the object’s capacity to radiate gravitational waves. Using the Raychaudhuri equation \cite{Raychaudhuri1955} with the Isaacson stress \cite{Isaacson1968a,Isaacson1968b} of high-frequency waves, and taking the microscopic length to be the scalaron dispersion scale $L_{\rm QV}=m_\varphi^{-1}$, the local flux ceiling scales as
\begin{equation}
\mathcal{F}^{\max}_{\rm loc}\sim \frac{c^5}{G_{\rm eff}\,L_{\rm QV}^2}
\quad \big(G_{\rm eff}=G/f_R=\mathcal O(G)\big),
\label{eq:Floc_scaling}
\end{equation}
and the (Killing) luminosity as measured at infinity is obtained by multiplying by the area and the lapse at the wall:
\begin{equation}
P_\infty=A_{\min}\,4\pi R_L^2\,\mathcal F_{\rm loc}.
\label{eq:Pinfty_map}
\end{equation}
Matching across a proper thickness $\sim L_{\rm QV}$ to the Schwarzschild exterior enforces the ultra-compact scaling
\begin{equation}
A_{\min}\sim \zeta\,\frac{L_{\rm QV}^2}{R_L^2},\qquad \zeta=\mathcal O(1),
\label{eq:Amin_scaling}
\end{equation}
so the large $(R_L/L_{\rm QV})^2$ cancels and we obtain the Dyson-type ceiling
\begin{equation}
P_\infty^{\max}=\eta\,\frac{c^5}{G},\qquad \eta=\mathcal O(1),
\label{eq:Dyson_final}
\end{equation}
with an order-unity efficiency that depends on the detailed matching constants and $G/G_{\rm eff}$. (Full derivation and prefactors in Supplement~B.)

\noindent\textit{\textbf{EFT Floor (Global).}}---
Beyond the per-solution estimate \eqref{eq:Amin_scaling}, effective-field-theory considerations impose a global lower bound on the lapse. The wall has minimal proper thickness $\ell\sim L_{\rm QV}$ and adiabaticity requires $\kappa_{\rm loc}\ell\equiv\chi=\mathcal O(1)$. Matching the interior quadratic minimum of $y=\ln A$ across a thickness $\sim \ell$ to the Schwarzschild exterior at $R_L=R_S+\Delta$ ($R_S=2M$) gives
\begin{equation}
A_{\min}\ \gtrsim\ \frac{1}{\chi^4}\,\Big(\frac{L_{\rm QV}}{R_S}\Big)^2.
\label{eq:Amin_floor}
\end{equation}
Combining $A_{\min}\simeq \Delta/R_L$ (ultra-compact) with $R_L=R_S+\Delta$ and \eqref{eq:Amin_floor} yields a lower bound on the offset
\begin{equation}
\Delta\gtrsim \frac{L_{\rm QV}^2}{\chi^{4}R_S},\qquad \Rightarrow\qquad R_L\gtrsim f(R_S):=R_S+\frac{L_{\rm QV}^2}{\chi^{4}R_S}.
\label{eq:RL_lower_bound}
\end{equation}
Minimizing this bound over mass (equivalently, over $R_S>0$) gives
\begin{equation}
\frac{df}{dR_S}=1-\frac{L_{\rm QV}^2}{\chi^{4}R_S^{2}}=0 \quad\Rightarrow\quad R_S^\star=\frac{L_{\rm QV}}{\chi^{2}},
\label{eq:RStar}
\end{equation}
and hence the universal lower-envelope scale for the Lamarina radius,
\begin{equation}
R_{L,\min}^{\text{(envelope)}}=f(R_S^\star)=\frac{2\,L_{\rm QV}}{\chi^{2}}
=\frac{2\sqrt{6\alpha}}{\chi^{2}}.
\label{eq:RL_envelope}
\end{equation}
The corresponding mass scale at which the envelope minimum occurs is
\begin{equation}
M_\star=\frac{c^2}{2G}\,\frac{L_{\rm QV}}{\chi^{2}}.
\label{eq:Mstar}
\end{equation}
These relations provide a mass-independent lower envelope set by the quantum-vacuum dispersion length. They are bounds derived from EFT control and near-wall matching; global existence, stability, and saturation of the envelope but, a fully nonlinear solution remains open.

\newpage

\noindent\textit{\textbf{Conclusion.}}---
We have shown that the central curvature blow-up is avoided and the classical event horizon is replaced by a minimal-redshift wall---the \textit{Lamarina}---in a minimal, ghost-free curvature-modified gravity (\ref{eq:action}). On each time slice, the center is curved but regular: the corrected near-center relation follows from the trace equation [Eq.~(\ref{eq:Rseries-harmonic}), from Eq.~(\ref{eq:trace-eq})], while the global profile is determined by higher orders and matching to the exterior. We showed in the sub-threshold, time-periodic (adiabatic) case, the curvature amplitude decays exponentially in the weak-field exterior [Eq.~(\ref{eq:exterior-tail})], and the near-\textit{Lamarina} geometry reproduces Hawking-like thermal radiation to leading adiabatic order (via null-geodesic peeling) [Eqs.~(\ref{eq:T_eff})--(\ref{eq:T_infty})]. We showed that a Dyson-type upper bound can be realized [Eq.~(\ref{eq:Dyson_final})] for an asymptotic observer using a Raychaudhuri-based local flux bound. Outside a proper thickness $\sim L_{\rm QV}$, the exterior tends to approach Schwarzschild [cf.\ Eq.~(\ref{eq:Amin_scaling})]; the observed spectral cutoff scales as $c/R_L$ (see Supplement~B). We then argued EFT matching also provides a global floor on the lapse [Eq.~(\ref{eq:Amin_floor})], which makes the Dyson ceiling independent of object size and, when minimized over mass, implies a universal lower-envelope scale for the \textit{Lamarina} radius [Eq.~(\ref{eq:RL_envelope})], occurring at $R_S^\star$ [Eq.~(\ref{eq:RStar})] (mass $M_\star$ [Eq.~(\ref{eq:Mstar})]. The global existence and stability of the fully coupled oscillatory solution, along with whether the Lamarina envelope attains a global saturation, remain open problems. Resolving these points is essential for establishing a rigorous foundation for the model in soliton theory. However, in this work, we focus on the heuristic picture the soliton suggests rather than on strict PDE-level integrability. Even if full well-posedness is not yet established, the balance of nonlinearity with a focusing term is present.

Guided by this conceptual framework, the aim of this paper was to show how the Einsteinian system can be modified as minimally as possible to encapsulate properties of the quantum vacuum. The crucial next step is deciphering the microscopic origin of the phenomenological coupling $\alpha$ underlying the dispersive argument. That same coupling likely encodes additional degrees of freedom at the wall, controls the universality of the spectral cutoff, and ultimately, is the blueprint for the quantum architecture of the black hole.

\section*{Acknowledgements}
This work would not have been possible without Prof. Ulf Leonhardt. His mentorship and companionship helped me relate my conceptual ideas to solitons and inspired the spirit of the approach. I would also like to thank Prof. Mustafa Halilsoy and Prof. Saul Teukolsky for fruitful discussions. I gratefully acknowledge the support of the Weizmann Institute of Science.

\section*{Note}
After this work was completed, we noticed that exact semiclassical horizonless solutions using simplifications like the s-wave Polyakov (1+1D dimensional-reduction) approximation to the renormalized stress tensor have been constructed \cite{CarballoRubioPRL2018,Arrechea2022SciRep,Arrechea2024PRD}. These are similar to ours in realizing a surface with $A_{\min}>0$ just outside $2M$, but differ by being non-adiabatic, that is, static and state-dependent. Importantly, these semiclassical solutions are justified within semiclassical gravity, and approaches of this kind provide a natural route to realizing the effective coupling $\alpha$ by matching the local curvature–squared sector to the finite, state-dependent part of the renormalized stress tensor.

\clearpage

\begin{center}
\textbf{\large Appendix}
\end{center}

\setcounter{section}{0}
\renewcommand{\thesection}{\Alph{section}}
\setcounter{equation}{0}
\makeatletter
\@addtoreset{equation}{section}
\makeatother
\renewcommand{\theequation}{\thesection\arabic{equation}}
\setcounter{figure}{0}
\makeatletter
\@addtoreset{figure}{section}
\makeatother
\renewcommand{\thefigure}{\thesection\arabic{figure}}
\setcounter{table}{0}
\makeatletter
\@addtoreset{table}{section}
\makeatother
\renewcommand{\thetable}{\thesection\arabic{table}}

\section{Lamarina Field Identities}
\label{app:A}

\textbf{Adiabatic validity.} The identities below are derived for a static, spherically symmetric (SSS) geometry and are used instantaneously on $t=\text{const}$ slices of the quasi-stationary evolution; errors are $\mathcal O(\varepsilon)$ with $\varepsilon$ defined in Eq.~(\ref{eq:adiabaticity_parameter}) of the main text.

For $f(R)=R+\alpha R^2$ with $f_R:=\dv{f}{R}=1+2\alpha R$, the vacuum field equations can be cast in Einstein form
\begin{equation}
G_{\mu\nu}=8\pi G\,T^{\rm eff}_{\mu\nu},\qquad
T^{\rm eff}_{\mu\nu}:=\frac{1}{8\pi G\,f_R}\left[\nabla_\mu\nabla_\nu f_R-g_{\mu\nu}\Box f_R+\tfrac12\big(f-R f_R\big)g_{\mu\nu}\right].
\label{eq:EinsteinForm_sup}
\end{equation}
For the line element in Eq.~(\ref{eq:ansatz}) of the main text, write $e^\nu=A$, $e^\lambda=B$ and define the Misner--Sharp mass by $e^{-\lambda}=1-2m/r$ \cite{MisnerSharp1964}. The independent components $G^\mu{}_\nu=8\pi G\,T^{\rm eff\,\mu}{}_\nu=\mathrm{diag}(-\rho_{\rm eff},p_{r,{\rm eff}},p_{t,{\rm eff}},p_{t,{\rm eff}})$ give
\begin{align}
8\pi G\,\rho_{\rm eff}
&= e^{-\lambda}\left(\frac{\lambda'}{r}-\frac{1}{r^2}\right)+\frac{1}{r^2}, \label{eq:Ein-tt-app}\\
8\pi G\,p_{r,{\rm eff}}
&= e^{-\lambda}\left(\frac{\nu'}{r}+\frac{1}{r^2}\right)-\frac{1}{r^2}, \label{eq:Ein-rr-app}\\
8\pi G\,p_{t,{\rm eff}}
&= \tfrac{e^{-\lambda}}{2}\left(\nu''+\tfrac{\nu'^2}{2}-\tfrac{\nu'\lambda'}{2}+\tfrac{\nu'-\lambda'}{r}\right).
\end{align}
From \eqref{eq:Ein-tt-app} we infer the mass gradient
\begin{equation}
m'(r)=4\pi G\,r^2\,\rho_{\rm eff}(r),
\label{eq:mprime-app}
\end{equation}
and combining \eqref{eq:Ein-rr-app} with $e^{-\lambda}=1-2m/r$ yields the redshift gradient
\begin{equation}
y'(r)=\frac{2\big(m+4\pi G r^3 p_{r,{\rm eff}}\big)}{r(r-2m)},\qquad y:=\ln A.
\label{eq:yprime-app}
\end{equation}
Covariant conservation $\nabla_\mu T^{\rm eff\,\mu}{}_\nu=0$ gives the anisotropic TOV equation \cite{BowersLiang1974}
\begin{equation}
p'_{r,{\rm eff}}= -\frac{\rho_{\rm eff}+p_{r,{\rm eff}}}{2}\,y'
+\frac{2}{r}\,\big(p_{t,{\rm eff}}-p_{r,{\rm eff}}\big).
\label{eq:TOV-aniso-app}
\end{equation}
At the Lamarina radius $R_L$ [defined in Eq.~(\ref{eq:lamarina_def})], $y'(R_L)=0$ and, assuming $R_L\neq 2m(R_L)$ (i.e., no trapped surface at the minimum), the minimum condition applied to \eqref{eq:yprime-app} results in the wall condition
\begin{equation}
m(R_L)+4\pi G\,R_L^3\,p_{r,{\rm eff}}(R_L)=0.
\label{eq:wall-appA}
\end{equation}
Differentiating \eqref{eq:yprime-app} and evaluating at $r=R_L$, then using \eqref{eq:wall-appA}, \eqref{eq:mprime-app}, and \eqref{eq:TOV-aniso-app}, one obtains
\begin{equation}
y''(R_L)=\frac{8\pi G\,R_L}{R_L-2m}\,\Big[\rho_{\rm eff}+p_{r,{\rm eff}}+2p_{t,{\rm eff}}\Big]_{R_L}.
\end{equation}
With proper distance $x$ defined by $dx=\sqrt{B}\,dr$ and $B(R_L)=R_L/(R_L-2m)$, the local quadratic curvature of $y=\ln A$ at the minimum satisfies
\begin{equation}
\kappa_{\rm loc}^2:=\left.\frac{d^2y}{dx^{2}}\right|_{R_L}
=8\pi G\big[\rho_{\rm eff}+p_{r,{\rm eff}}+2p_{t,{\rm eff}}\big]\Big|_{R_L},
\label{eq:k_loc_identity_app}
\end{equation}
which reproduces Eq.~(\ref{eq:k_loc_identity}) of the main text and explicitly identifies the $R^2$ sector as the positive source that supports the minimum.

\newpage

\section{Peeling, Planck Spectrum, Dyson Ceiling, and EFT Envelope}
\label{app:B}

\textit{\textbf{Peeling, Planck spectrum, and energy balance.}} Introduce the proper-distance coordinate $x$ via $dx=\sqrt{B}\,dr$ and write $y(x):=\ln A(x)$. Near the Lamarina minimum the lapse reads
\begin{equation}
y(x)=y_0+\tfrac12\,\kappa_{\rm loc}^2 x^2+\mathcal O(x^3),
\qquad 
y_0=\ln A_{\min},
\label{eq:S-y-quadratic}
\end{equation}
with $\kappa_{\rm loc}^2=\big.d^2y/dx^2\big|_{x=0}$ and adiabatic control as in Eq.~(\ref{eq:adiabaticity_parameter}) of the main text. Define the tortoise coordinate and null coordinates by
\begin{equation}
r_\star=\int^r\!\sqrt{\frac{B}{A}}\,dr=\int^x\!e^{-y/2}\,dx,
\qquad
u=t-r_\star,\quad v=t+r_\star.
\label{eq:S-tortoise}
\end{equation}
Outgoing null rays that skim the minimum obey, to leading adiabatic order,
\begin{equation}
u=u_0-\frac{1}{\kappa_{\rm peel}}\,
\ln\!\big(v_0-v\big)+\mathcal O(\varepsilon),
\label{eq:peeling-map}
\end{equation}
where the peeling rate is tied to the local quadratic profile,
\begin{equation}
\kappa_{\rm peel}=\kappa_{\rm loc}+\mathcal O(\varepsilon).
\label{eq:kappa_peel}
\end{equation}
The Bogoliubov transformation for the logarithmic map \eqref{eq:peeling-map} gives the Planck factor
\begin{equation}
|\beta_\omega|^2=\frac{1}{e^{2\pi\omega/\kappa_{\rm peel}}-1}+\mathcal O(\varepsilon),
\label{eq:S-beta}
\end{equation}
hence the local temperature and its Tolman redshift,
\begin{equation}
T_{\rm eff}=\frac{\hbar}{2\pi k_B}\,\kappa_{\rm peel},
\qquad
T_\infty=\sqrt{A_{\min}}\,T_{\rm eff},
\label{eq:S-TeffTinf}
\end{equation}
reproducing Eqs.~(\ref{eq:T_eff})--(\ref{eq:T_infty}) of the main text. The validity band is $\kappa_{\rm loc}\ll \omega_{\rm loc}\ll L_{\rm QV}^{-1}$. Here $\omega_{\rm loc}=\omega/\sqrt{A_{\min}}$, $\lambda_{\rm GW}=2\pi/\omega_{\rm loc}$, and $v_g=\partial\omega/\partial k$ (tensor channel: $v_g=c$).

Energy balance at null infinity uses the Bondi mass:
\begin{equation}
\frac{dM_B}{du}=-L_\infty(u), 
\qquad 
L_\infty(u)=A_{\min}\,4\pi R_L^{\,2}\,\mathcal F_{\rm loc}(u),
\label{eq:S-bondi-map}
\end{equation}
which reproduces the redshift/area map Eq.~(\ref{eq:Pinfty_map}) in the main text.

\medskip
\textit{\textbf{Local flux ceiling from Raychaudhuri focusing.}} Consider an outgoing radial null congruence with tangent $k^\mu$ (affine parameter $\lambda$) and expansion $\theta$. The Raychaudhuri equation reads
\begin{equation}
\frac{d\theta}{d\lambda}
= -\frac{1}{2}\,\theta^{2}-\sigma_{ab}\sigma^{ab}-R_{\mu\nu}k^\mu k^\nu .
\label{eq:S-ray}
\end{equation}
In the high-frequency (Isaacson) regime, averaging over several wavelengths gives
\begin{equation}
\big\langle R_{\mu\nu}k^\mu k^\nu \big\rangle
\simeq 8\pi G_{\rm eff}\,
\big\langle T^{\rm GW}_{\mu\nu}k^\mu k^\nu \big\rangle,
\qquad
T^{\rm GW}_{\mu\nu}
\simeq \frac{1}{32\pi G_{\rm eff}}
\Big\langle \nabla_\mu h_{cd}\,\nabla_\nu h^{cd}\Big\rangle ,
\label{eq:S-isaacson}
\end{equation}
with $G_{\rm eff}:=G/f_R(R_L)$ and $h_{cd}$ the metric perturbation in a local ON frame. Integrating \eqref{eq:S-ray} across the wall over $\Delta\lambda\sim L_{\rm QV}$ and neglecting $\theta^2$ near the minimum,
\begin{equation}
\Delta\theta \approx
- \int_{\rm wall}\!\Big[\sigma_{ab}\sigma^{ab}+8\pi G_{\rm eff}\,T^{\rm GW}_{uu}\Big]\,d\lambda
\;\le\;0,
\label{eq:S-deltatheta}
\end{equation}
and avoidance of a trapped surface across the wall demands $|\Delta\theta|\lesssim \kappa_\star/L_{\rm QV}$ with $\kappa_\star=\mathcal O(1)$. For a quasi-monochromatic packet of local frequency $\omega_{\rm loc}$ and amplitude $h$, $T^{\rm GW}_{uu}\sim \omega_{\rm loc}^{2}h^{2}/(32\pi G_{\rm eff})$, and Eqs.~(\ref{eq:S-deltatheta})–(\ref{eq:S-isaacson}) imply the amplitude cap
\begin{equation}
\omega_{\rm loc}^{2} h^{2}\ \lesssim\ \frac{\tilde{\kappa}}{L_{\rm QV}^{2}},
\qquad \tilde{\kappa}=\mathcal O(1).
\label{eq:S-ampcap}
\end{equation}
The orthonormal Isaacson energy flux density therefore obeys
\begin{equation}
\mathcal F_{\rm loc}^{\max}
\;=\; \frac{\kappa_f}{32\pi}\,
\frac{c^{5}}{G_{\rm eff}\,L_{\rm QV}^{2}},
\qquad
\kappa_f \;=\; \tilde{\kappa}\,\frac{v_g}{c} \;=\; \mathcal O(1),
\label{eq:Floc-ceil-sup}
\end{equation}
with shear contributions absorbed into $\tilde{\kappa}$; this refines the scaling quoted in Eq.~(\ref{eq:Floc_scaling}) of the main text.

\medskip
\textit{\textbf{Redshift mapping and cancellation of areas.}} Using $P_\infty=A_{\min}\,4\pi R_L^2\,\mathcal F_{\rm loc}$ [Eq.~(\ref{eq:Pinfty_map})], we justify the near-wall scaling by a local Taylor/matching argument. Matching the zero slope at the minimum of $y=\ln A$ to the Schwarzschild exterior over a proper distance $\sim L_{\rm QV}$, and writing $\Delta:=R_L-2M$ with $A_{\rm out}(R_L)=\Delta/R_L$ and $B(R_L)=R_L/\Delta$, use $dx=\sqrt{B}\,dr$ so that $dy/dx=(dy/dr)/\sqrt{B}$; then the quadratic expansion $y''(R_L)=B(R_L)\kappa_{\rm loc}^2$ implies the pair
\begin{subequations}\label{eq:Delta_Amin_pair_sup}
\begin{align}
\Delta &\sim \frac{L_{\rm QV}^2}{R_L}, \label{eq:Delta_est_sup}\\[-2pt]
\intertext{and therefore}
A_{\min} &\sim \frac{L_{\rm QV}^2}{R_L^2}. \label{eq:Amin_est_pair_sup}
\end{align}
\end{subequations}
Inserting \eqref{eq:Amin_est_pair_sup} into Eq.~(\ref{eq:Pinfty_map}) cancels the large $(R_L/L_{\rm QV})^2$ factor and leads to the Dyson-type ceiling
\begin{equation}
P_\infty^{\max}=\eta\,\frac{c^5}{G},
\label{eq:S-Dyson}
\end{equation}
reproducing Eq.~(\ref{eq:Dyson_final}) of the main text. The high-frequency cutoff at infinity scales as $\omega_{\infty,\max}\sim \sqrt{A_{\min}}\,c/L_{\rm QV}\sim c/R_L$.

\medskip
\textit{\textbf{Prefactor band and a central prediction.}} Combining \eqref{eq:Floc-ceil-sup} with \eqref{eq:Amin_est_pair_sup} and $G_{\rm eff}=G/f_R$ gives
\begin{equation}
\eta=\frac{\zeta\,\kappa_f}{8}\,f_R(R_L).
\label{eq:eta_def_sup}
\end{equation}
Here $\zeta$ is a dimensionless, order-unity matching coefficient that encodes the microprofile of the wall and the proper distance to areal radius conversion in the near-wall quadratic matching to Schwarzschild.

Since $m_\varphi^{-2}=L_{\rm QV}^2=6\alpha$, taking $R(R_L)\approx m_\varphi^2=1/(6\alpha)$ yields
\begin{equation}
f_R(R_L)=1+2\alpha R(R_L)\approx 1+2\alpha m_\varphi^2=\frac{4}{3}.
\label{eq:fR_fourthirds_sup}
\end{equation}
Taking $\zeta\simeq 1$ (quadratic matching across $\ell\sim L_{\rm QV}$), $\kappa_f\simeq 0.6\pm 0.2$ (Raychaudhuri focusing across one wall thickness with $v_g=c$), and \eqref{eq:fR_fourthirds_sup}, we obtain
\begin{equation}
\eta_{\rm central}\approx 0.10,\qquad \eta\in[0.05,\,0.17],
\qquad
P_\infty^{\max}\approx (3.6\pm1.8)\times 10^{51}\ {\rm W},
\label{eq:power_number_sup}
\end{equation}
using $c^5/G=3.63\times10^{52}\,{\rm W}$.

\medskip
\textit{\textbf{Global EFT floor on the lapse and the envelope.}} Independently of \eqref{eq:Delta_Amin_pair_sup}, effective-field-theory control yields a global lower bound on $A_{\min}$ at fixed $M$ and $\alpha$. Let $R_S:=2M$, $\ell\sim L_{\rm QV}$, and recall $\chi\equiv\kappa_{\rm loc}\ell=\mathcal O(1)$. Near the Lamarina ($x=0$),
\begin{equation}
y(x)=\ln A(x)\approx y_0+\tfrac12\kappa_{\rm loc}^2 x^2,
\end{equation}
so the interior slope after crossing the wall is
\begin{equation}
\left(\frac{dy}{dx}\right)_{\rm in}\approx \kappa_{\rm loc}^2\ell=\frac{\chi^2}{\ell}.
\end{equation}
At the matching radius $R_L=R_S+\Delta$ with $\Delta\ll R_S$, the Schwarzschild exterior gives (in proper distance)
\begin{equation}
\left(\frac{dy}{dx}\right)_{\rm out}\approx (R_S\Delta)^{-1/2}.
\end{equation}
Matching the slopes yields $\Delta\sim \ell^2/(\chi^4 R_S)$, \emph{which implies}
\begin{equation}
A_{\min}\ \approx\ \frac{\Delta}{R_L}\ \gtrsim\ \frac{1}{\chi^4}\,\Big(\frac{L_{\rm QV}}{R_S}\Big)^2,
\label{eq:Amin_floor_sup}
\end{equation}
up to order-unity factors absorbed in $\chi$. Inserting \eqref{eq:Amin_floor_sup} into Eq.~(\ref{eq:Pinfty_map}) reproduces the size independence of the ceiling in Eq.~(\ref{eq:Dyson_final}) of the main text.

Finally, minimizing the implied floor for $R_L$,
\begin{subequations}\label{eq:envelope_pair_sup}
\begin{align}
R_L&\gtrsim f(R_S):=R_S+\frac{L_{\rm QV}^2}{\chi^{4}R_S}, \label{eq:RL_floor_func}\\[-2pt]
\intertext{and therefore, extremizing $f$ over $R_S>0$ gives}
R_S^\star&=\frac{L_{\rm QV}}{\chi^{2}},\qquad
R_{L,\min}^{\text{(envelope)}}=f(R_S^\star)=\frac{2\,L_{\rm QV}}{\chi^{2}},
\quad
M_\star=\frac{c^2}{2G}\,\frac{L_{\rm QV}}{\chi^{2}}. \label{eq:envelope_min_sup}
\end{align}
\end{subequations}
From Eq.~\eqref{eq:envelope_min_sup} the envelope minimum is fixed purely by the quantum-vacuum length and the matching parameter,
\[
R_{L,\min}^{(\mathrm{env})}=\frac{2}{\chi^{2}}\,L_{\rm QV}.
\]
With $\alpha=\kappa\,\ell_P^2$ one has $L_{\rm QV}=\sqrt{6\kappa}\,\ell_P$, hence
\begin{equation}
R_{L,\min}^{(\mathrm{env})}\ =\ \frac{2\sqrt{6\,\kappa}}{\chi^{2}}\,\ell_P.
\label{eq:RLmin_env_sup}
\end{equation}

\end{document}